RESEARCH ARTICLE

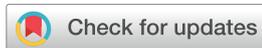

# Expanding the tunability and applicability of exchange-coupled/decoupled magnetic nanocomposites†

Cecilia Granados-Miralles, [ID] *[a] Adrián Quesada, [ID] [a] Matilde Saura-Múzquiz, [ID] ‡[b] Henrik L. Andersen, [ID] §[b] José F. Fernández [ID] [a] and Mogens Christensen [ID] [b] $CoFe_2O_4$/Co–Fe magnetic composites are usually prepared through partial reduction of $CoFe_2O_4$, which often yields monoxides (*i.e.*, FeO, CoO) as secondary phases. Since these compounds are paramagnetic at ambient conditions, the presence of a small amount of monoxide is generally downplayed in the literature, and the possible effects on the magnetic properties are simply ignored. However, the present study shows that even a low concentration of monoxide results in decoupling of the soft and hard magnetic phases, which inevitably leads to a deterioration of the magnetic properties. Additionally, it is confirmed that a partial reduction of $CoFe_2O_4$ is a suitable method to produce $CoFe_2O_4$/Co–Fe nanocomposites, provided that the treatment is well controlled with respect to duration, temperature and flow of reductant. A monoxide-free nanocomposite was produced and its magnetic properties evaluated both at room and low temperature. Our model system exemplifies the potential of exchange-coupling (and decoupling) as a tool to tune the magnetic properties of a material within a relatively wide range of values, thus widening its spectrum of potential applications.

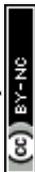

## Introduction

Magnetic nanoparticles (MNPs) have undoubtedly been one of the hot research topics of the 21st century.[1] Intensive research on the subject has yielded notable advances in a wide range of technologies and disciplines. For instance, MNPs have been a great aid in medical diagnosis and treatment of diseases.[2] Among other cutting-edge medical applications, MNPs are integral components of drug carriers for magnetic drug delivery,[3,4] heat mediators in cancer therapy by magnetic fluid hyperthermia (MFH),[5] or contrast agents for magnetic resonance imaging (MRI).[6] MNPs are also highly relevant in matter of sensors and biosensors aimed to diverse analytes,[7] *e.g.*, food contaminants,[8,9] environmental pollutants,[10] antibodies,[11] *etc.*

The actual application determines the required magnetic properties. Very often, the stability and longevity of the devices rely on a strong resistance to demagnetization (*i.e.* hard magnetic material, with large coercivity, $H_c$). Other times, the crucial parameter that ensures compliance with the specific task is the ability of the material to become magnetized up to a high value (*i.e.* high saturation magnetization, $M_s$). Most of the available materials show either a large $H_c$ and a moderate $M_s$ or *vice versa*.[12] Consequently, if relatively high values of both $H_c$ and $M_s$ are necessary, fabrication of composite materials should be addressed. According to the exchange-spring theory, the $M_s$ of a hard magnetic material can be enhanced by adding a controlled amount of a large-$M_s$ material (generally soft), and the cost in $H_c$ will be low provided that the two materials are effectively exchange-coupled.[13]

Ferrites are among the most used magnetic materials, owing to their good magnetic properties, chemical and mechanical stability, and the availability of elements they are based on. Especially interesting are the spinel ferrites (SFs), as they allow easy tunability of the magnetic properties with small changes on the chemical composition,[14–16] thus increasing their versatility towards different applications. SFs have been widely used in the electronic industry, for high-density data storage and spintronic devices.[17,18] Their utilization for biomedical applications has increased significantly over the last years, especially in the fields of drug delivery[19] and biosensors.[20,21] In addition to their applications as magnetic materials, it is worth mentioning that SFs are widely used for other purposes, *e.g.*, as catalysts for very varied chemical processes,[22,23] advanced battery electrodes,[24,25] electrochemical supercapacitors in energy storage systems,[26] *etc.*

[a] *Electroceramics Department, Instituto de Cerámica y Vidrio, CSIC, Kelsen 5, ES 28049 Madrid, Spain. E-mail: c.granados.miralles@icv.csic.es*

[b] *Center for Materials Crystallography, Department of Chemistry and iNANO, Aarhus University, Langelandsgade 140, DK 8000 Aarhus, Denmark.*

† Electronic supplementary information (ESI) available. See DOI: 10.1039/c9qm00713j

‡ Current address: School of Chemistry, University of Sydney, F11, NSW 2006 Sydney, Australia.

§ Current address: School of Chemistry, UNSW Australia, NSW 2052 Sydney, Australia.

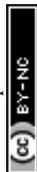

SFs have the general formula $M^{2+}(Fe^{3+})_2O_4$, with M = Mg, Mn, Fe, Co, Ni, Cu, Zn.[17] Out of all them, only Co-spinel shows hard magnetic properties, while the rest are soft magnetic species.[27] Moreover, $CoFe_2O_4$ can be easily reduced to a Co–Fe alloy in the presence of a small concentration of $H_2$ gas and moderate temperatures ($\approx 300\ ^\circ C$).[28] Both facts make this compound interesting, as an incomplete $CoFe_2O_4$ reduction directly leads to coexistence of hard ($CoFe_2O_4$) and soft (Co–Fe) magnetic phases. This is an excellent tool from the material science viewpoint, as it offers the potential to fine tuning the soft/hard magnetic behavior of the produced material by means of controlling the composite composition.

For the above reasons, numerous studies on the $CoFe_2O_4$ (hard)/Co–Fe (soft) composite are found in the literature, including composites prepared as powders,[29] dense pellets,[30] or thin films.[31] Some works have set the main focus on the preparation process (*in situ* studies),[28,32] while others have taken care of an in-depth structural characterization of the produced composites using spectroscopic techniques such as Raman[33] or Mössbauer spectroscopy.[34,35] Others have put great efforts on studying the inter-particle coupling from different perspectives, both using transmission electron microscopy (TEM), and measuring $\delta m$ curves (Henkel plots).[35,36] Recently, micromagnetic calculations on these systems have also been reported.[37] However, a successful exchange-coupling of these two magnetic phases has proven rather challenging to achieve, the reason behind it often remaining unclear. In the present work, the origin of magnetic decoupling in $CoFe_2O_4$/Co–Fe nanocomposites is addressed. Composites covering a range of compositions are prepared, and their crystalline and atomic structures are studied using high-resolution powder X-ray diffraction. Physical characterization of the magnetic properties is carried out both at room and low temperature, and coupling/decoupling of the system is evaluated in terms of the phases present in the sample and their average crystallite sizes.

## Experimental

### Sample preparation

Magnetic $CoFe_2O_4$/Co–Fe nanocomposites were prepared by means of a controlled reduction of $CoFe_2O_4$ nanoparticles. The starting $CoFe_2O_4$ material was hydrothermally synthesized following the procedure described in a previous work,[38] and had a volume-averaged crystallite size of 14.4(1) nm. 0.20 g of the as-synthesized powders were spread on an $Al_2O_3$ crucible with approximate dimensions $60 \times 40\ mm^2$. The crucible was placed at the hot spot of a tubular furnace (C.H.E.S.A. Ovens). The furnace was sealed at both ends and purged down to a pressure of $\approx 1 \times 10^{-2}$ mbar using a vacuum pump connected to the furnace outlet. Gas mixture 10% $H_2$/90% $N_2$ was fed through the furnace inlet, regulating the flow until the pressure inside the furnace stabilized at 20 mbar. Finally, the thermal treatment was initiated. An initial heating ramp of 100 $^\circ C\ min^{-1}$ drove the temperature up to the set point (300–600 $^\circ C$), at which the system was maintained during 2–8 hours (see heating profiles in Fig. S1, ESI†). Subsequently, the sample was left to cool down inside the furnace, while maintaining the flow of reducing gas. The sample was removed from the furnace once the temperature was below 75 $^\circ C$. All samples were stable in air.

### Characterization

**Powder X-ray diffraction (PXRD).** PXRD data were collected on all the samples in a Bragg-Brentano $\theta/\theta$ configuration using Cu K$\alpha_{1,2}$ radiation ($\lambda_1 = 1.540593$ Å, $\lambda_2 = 1.544427$ Å) at a laboratory Rigaku SmartLab® diffractometer operated at 40 kV and 180 mA. The incident slit (IS) choice was different depending on the amount of sample available for the measurement. Further details on IS and $2\theta$ range may be found in the ESI.† A diffracted beam monochromator (DBM) was installed on the receiving optics to suppress the fluorescence contribution to the background and the data were collected with a D/teX Ultra detector.

Rietveld analysis of the PXRD data was performed using the *FullProf* Suite.[39] In the Rietveld model, the oxides were described assuming a Co:Fe stoichiometry of 1:2 (*i.e.*, $CoFe_2O_4$, $Co_{0.33}Fe_{0.67}O$) and a random distribution of the two cations among the equivalent crystallographic sites. The elemental composition of the alloy in the model varied depending on the sample. A detailed crystallographic description of all the Rietveld phases may be found on Tables S1–S5 in the ESI.†

Data were also collected on a NIST 660B LaB$_6$ calibrant in the different experimental configurations, and these data were modelled (LeBail fit) to estimate the instrumental contribution to the peak broadening. The instrument contribution was deconvoluted from the samples data, and the remaining profile broadening, originating from the sample, was modelled as Lorentzian isotropic size-broadening using the Thompson–Cox–Hastings formulation of the pseudo-Voigt function.[40]

**Magnetic properties.** About 10 mg of the nano-powders, measured with a precision of 0.001 mg, were gently compressed into thin cylindrical pellets (diameter = 3.00 mm, thickness = 0.50–0.60 mm). Magnetization as a function of an externally applied magnetic field was measured using a Quantum Design Physical Property Measurement System (PPMS®) equipped with a vibrating sample magnetometer (VSM). After field-cooling in 50 kOe (*i.e.*, 3979 kA m$^{-1}$) down to 10 K, the magnetization was measured while varying the applied field in the range ±50 kOe. Subsequently, the sample was heated up to 300 K, and the magnetization was measured in the field range ±20 kOe (1591 kA m$^{-1}$). For the starting material, the LT measurement was done after cooling in absence of an external field.

Prior to the measurements described above, the room temperature magnetization of the samples was measured in a smaller field range ±4 kOe (318 kA m$^{-1}$) using a home-built VSM setup.[41]

## Results and discussion

### Composition and crystallite size from Rietveld analysis

Reduction treatments of variable duration and temperature yielded five different samples. Henceforth, tags in the form {time@temperature} are used to refer to the samples. Sample

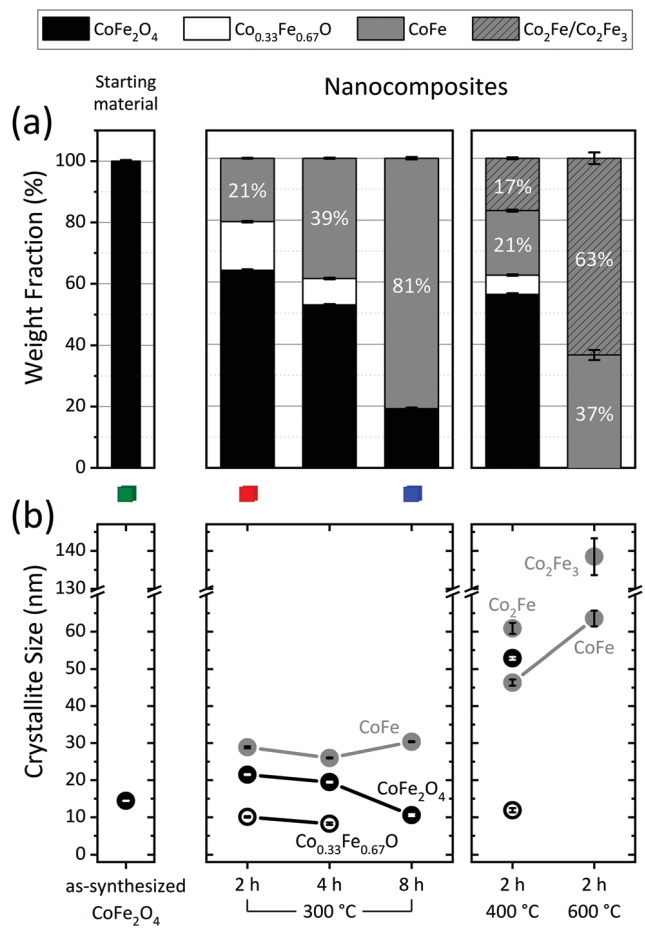
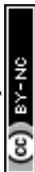

Fig. 1 (a) Sample composition and (b) crystallite size of the constituent phases extracted from Rietveld analysis of the PXRD data measured on the starting material and the five different nanocomposites. The magnetic properties of the samples highlighted with a green/red/blue square are represented in Fig. 4 and 5 using the same color-code.

composition and sizes obtained from Rietveld analysis of the PXRD data collected on those samples are displayed in Fig. 1 and Table 1. A representative example of a Rietveld model fitted to the PXRD data is shown in Fig. 2(a). The Rietveld models fitted to the PXRD data collected for the remaining samples may be found on Fig. S5 in the ESI.†

From the series of experiments at 300 °C with variable duration (2–8 h), it is clear that as time increases, the amount of $CoFe_2O_4$ decreases, at the expense of the appearance of reduced phases: a monoxide phase ($Co_{0.33}Fe_{0.67}O$) and a metallic alloy phase (CoFe). The monoxide seems to play the role of a reaction intermediate, as it disappears as the reduction advances. Thus, while 2 and 4 h at 300 °C produced composites with 16.1(2)% and 8.6(3)% monoxide, respectively, a monoxide-free composite with an 80.9(4)% metallic content was obtained after 8 h. Fig. 2(b–d) show selected 2θ-regions of the PXRD data and models corresponding to these three samples. The distinct Rietveld phases are highlighted to illustrate the appearance/disappearance of the different phases as dwell time increases.

At 300 °C, the growth of the soft phase crystallites remains relatively controlled (≤30.4(2) nm) regardless of the dwell time. Increasing the treatment temperature accelerates the reduction process,[28] thus, 2 h at 400 °C led to lower $CoFe_2O_4$ content than 2 h at 300 °C. The monoxide content also decreased substantially at 400 °C. At 600 °C, 2 hours were sufficient to completely reduce the starting material to pure metallic phases. However, increasing the temperature entails a significant growth of the alloy crystallites.

Fig. 3(a) shows the evolution of the most intense reflections of the alloy phase as a function of the reduction temperature. While the diffraction data collected for the {2h@300°C} nanocomposite can be modelled with a single metallic phase (CoFe), at least two metallic phases are clearly present in the {2h@400°C} and {2h@600°C} samples. The refined unit cell parameters for the individual phases are displayed in Table 1 and plotted in Fig. 3(b) as a function of the treatment temperature. The dissimilar distribution of cell parameters suggests different elemental compositions of the alloys. Unfortunately, the Co:Fe ratio could not be extracted from the refinements, because Co and Fe are next-neighbors in the periodic table and therefore, practically indistinguishable using X-rays (see ESI in ref. 28).

The unit cell dimensions of Co–Fe alloys increase with an increasing Fe content.[42] This allows an estimate of the elemental composition based on the lattice parameter. The empirical chemical compositions shown in Table 1 and Fig. 3 were assessed by substituting the refined unit cell parameters in the equation obtained by Ohnuma et al. for ordered body-centered-cubic

Table 1 Results from Rietveld refinements for the three phases, i.e., weight fraction, wt%, volume-averaged crystallite size, ⟨D⟩, unit cell parameter, a. The uncertainties specified along with the refined wt% and a values are the standard deviations calculated by the refinement software (FullProf).[39] For the refined ⟨D⟩, the uncertainties are calculated by propagation of error

| | $CoFe_2O_4$ | | | $Co_{0.33}Fe_{0.67}O$ | | | Metallic alloy | | | |
|---|---|---|---|---|---|---|---|---|---|---|
| Sample | wt% | ⟨D⟩ (nm) | a (Å) | wt% | ⟨D⟩ (nm) | a (Å) | wt% | | ⟨D⟩ (nm) | a (Å) |
| Starting material | 100.0(3) | 14.4(1) | 8.3929(1) | — | — | — | — | | — | — |
| {2h@300°C} | 63.8(3) | 21.5(1) | 8.3889(1) | 15.7(2) | 10.1(2) | 4.2695(2) | 20.5(1) | CoFe | 28.9(3) | 2.85645(4) |
| {4h@300°C} | 52.7(3) | 19.5(1) | 8.3886(1) | 8.5(2) | 8.3(3) | 4.2680(4) | 38.9(2) | CoFe | 26.0(2) | 2.85877(4) |
| {8h@300°C} | 19.1(4) | 10.6(3) | 8.3904(5) | — | — | — | 80.9(4) | CoFe | 30.4(2) | 2.86141(4) |
| {2h@400°C} | 56.1(4) | 52.9(4) | 8.38952(7) | 6.2(2) | 11.9(6) | 4.2860(4) | 37.8(7) | $Co_2Fe$ | 61(2) | 2.84639(3) |
| | | | | | | | 20.8(3) | CoFe | 46.3(8) | 2.85434(4) |
| {2h@600°C} | — | — | — | — | — | — | 37(2) | CoFe | 64(2) | 2.85954(6) |
| | | | | | | | 64(2) | $Co_2Fe_3$ | 139(5) | 2.86405(3) |

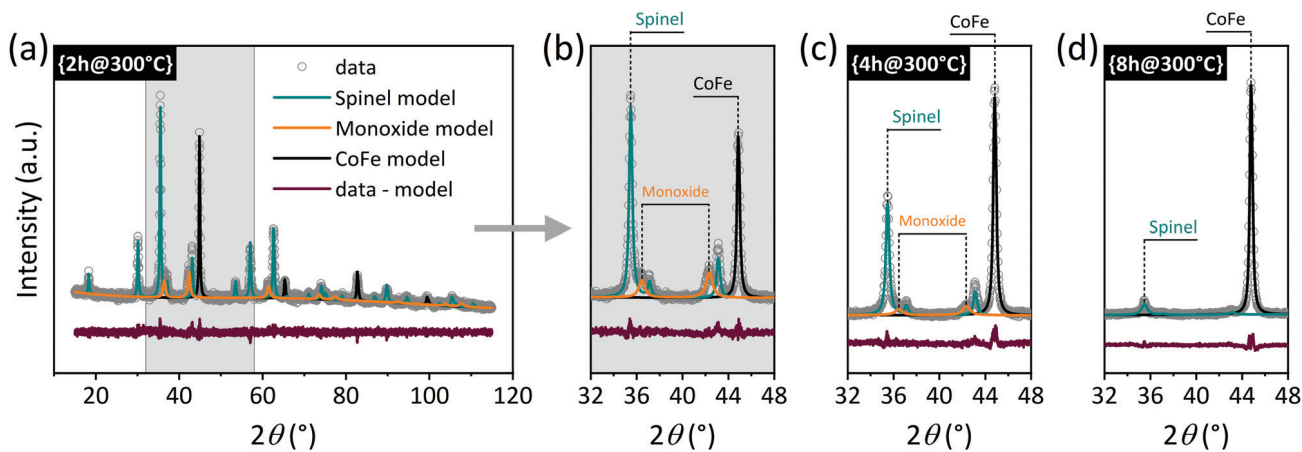

Fig. 2  (a) PXRD data and corresponding Rietveld model of the phases present in sample {2h@300°C}. (b) Selected 2θ-region of data and models for {2h@300°C}, (c) {4h@300°C}, and (d) {8h@300°C}.

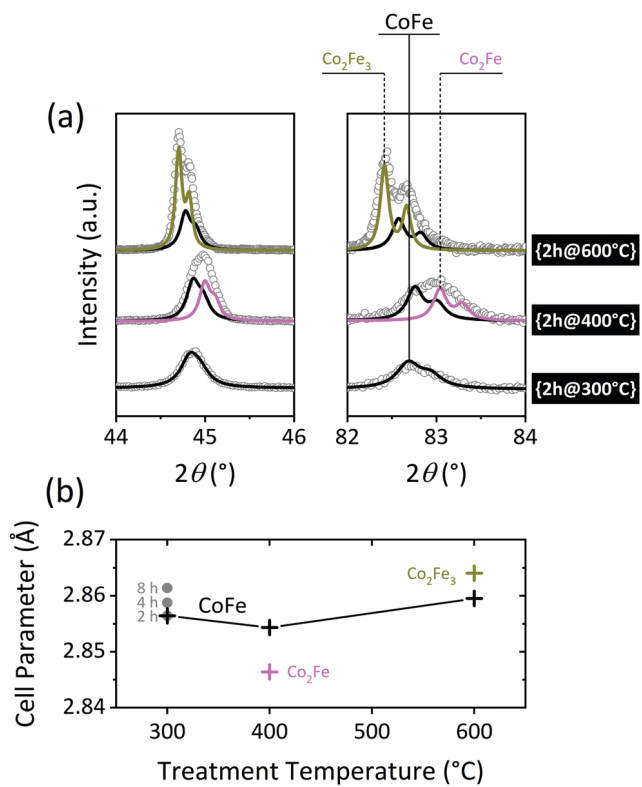

Fig. 3  (a) Selected 2θ-regions of the PXRD data collected after 2 h reduction treatments at 300, 400, and 600 °C, and Rietveld models of the different metallic phases, i.e., $Co_2Fe_3$, CoFe, and $Co_2Fe$. (b) Refined unit cell parameters of the phases as a function of the treatment temperature, circles and crosses representing the time and temperature series, respectively. The error bars lie within the size of the symbols.

(bcc) structures.[42] For the mildest reduction, {2h@300°C}, the calculated alloy composition is CoFe. This indicates surplus Co on the alloy, compared to the Co:Fe stoichiometry of 1:2 presumed for the starting spinel material. This observation is in agreement with previous *in situ* investigations on this system, where the reduced phases were observed to appear in a Co-rich form, to later incorporate Fe and evolve towards Co:Fe = 1:2.[28] At the higher temperatures, CoFe coexists with other alloy phases, *i.e.*, $Co_2Fe$ in {2h@400°C} and $Co_2Fe_3$ in {2h@600°C}, showing that the Fe-content increases as the temperature rises. A similar phase segregation may be occurring at 300 °C, although the effect remains hidden under the broader diffraction peaks derived from the smaller crystallite sizes at this temperature, and in that case, the refined unit cell parameter should be understood as the weighted average of all the phases present. The cell dimensions increase slightly with dwell time, again indicating a late incorporation of the Fe in the alloy structure.

The influence of the amount of $H_2$ inside the furnace was also investigated (see Fig. S6 in the ESI†). The gas pressure was increased up to 100 and 300 mbar, and no significant changes were observed neither on the sample composition nor the crystallite sizes, compared to the experiments at 20 mbar. This suggests that, for the amounts of sample used here, an $H_2$ excess is ensured even at the lowest pressure, and as long as there is enough $H_2$ available, the gas pressure does not seem to have a major influence on the process.

To evaluate whether the crystallite size of the starting material plays a role, an additional time series of experiments were carried out at 300 °C using $CoFe_2O_4$ powders with an average size of 8.2(1) nm (see Fig. S7 in the ESI†). Comparing these results with those represented in Fig. 1 (mean size starting material 14.4(1) nm), it is concluded that the smaller the size of the starting $CoFe_2O_4$, the faster the reduction occurs, *i.e.*, the shorter the time required to achieve a certain reduction stage.

### Magnetic properties

**Magnetization at room temperature (RT).** Magnetic hysteresis loops measured at 300 K are displayed in Fig. S8 (ESI†) and saturation magnetization, $M_s$, remanence, $M_r$, and coercivity, $H_c$, obtained from those curves are compiled in Table 2 and plotted in Fig. 4 as a function of the alloy content. $M_s$ was calculated from the loops using the law of approach to saturation.[43] $M_r$ and $H_c$ were extracted from linear fits including 5 data points on each side of the *y*- and the *x*-intercept, respectively.

Table 2 Saturation magnetization, $M_s$, remanence, $M_r$, and coercivity, $H_c$, extracted from magnetic hysteresis measured at 300 K. The errors on the values are calculated from the uncertainties on the linear fits

| Sample | $M_s$ (A m² kg⁻¹) | $M_r$ (A m² kg⁻¹) | $H_c$ (kA m⁻¹)[a] | $H_c$ (kOe)[a] |
|---|---|---|---|---|
| Starting material | 73.9(4) | 19.7(1) | 83(2) | 1.04(2) |
| {2h@300°C} | 86.3(1) | 29.5(1) | 115(1) | 1.44(2) |
| {4h@300°C} | 115.6(1) | 30.4(2) | 90(1) | 1.13(2) |
| {8h@300°C} | 185.1(1) | 27.0(2) | 60.4(9) | 0.76(1) |
| {2h@400°C} | 125.6(1) | 13.8(2) | 44.3(6) | 0.557(7) |
| {2h@600°C} | 229.7(2) | 1.7(2) | 3.23(2) | 0.0406(2) |

[a] $H_c$ is given both in SI an CGS units to ease comparison with other works.

In order to discriminate the influence of the temperature from the effect of the actual reduction process, a 2 h long treatment in vacuum at 400 °C was carried out. No significant changes were observed in the magnetic properties after this treatment (see solid, gray circles in Fig. 4). Therefore, in the following, the starting CoFe₂O₄ powders will continue to be used as reference to evaluate the magnetic properties of the nanocomposites.

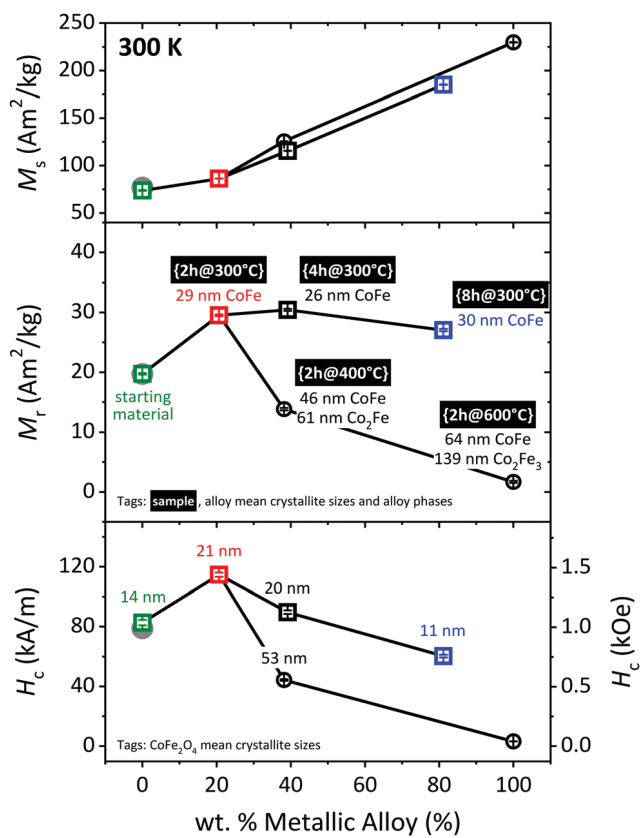

Fig. 4 Room temperature $M_s$, $M_r$, and $H_c$ as a function of the weight fraction of metallic alloy. The green, open squares correspond to the starting material, the rest of the squares represent the time series of experiments (at 300 °C), and the open circles the two high-temperature experiments (400 and 600 °C). The crystallite sizes indicated in the figure are relevant for the discussion of results in the text. The gray, solid circles correspond to a reference/blank sample fabricated from the same starting material, in a 2 h-long treatment in vacuum at 400 °C. The drawn lines are intended as a guide to the eye.

$M_s$ follows the expected linear increase with the amount of alloy present in the sample. The trends exhibited by $M_r$ and $H_c$ are slightly more complex. A mild reduction, such as {2h@300°C} (in red color) yields a significant enhancement of both parameters; the composite with a 20.5(1) wt% alloy has a 50% higher $M_r$ and a 39% larger $H_c$ than the starting material. This is understood as a consequence of the temperature which causes a moderate growth of the CoFe₂O₄ nanoparticles, from 14.4(1) to 21.5(1) nm, and has very likely induced a betterment of the crystallinity as well.

As the alloy wt% increases, both $M_r$ and $H_c$ decrease, but the decrease is much more pronounced for the temperature series (circles) than for the time series (squares). For instance, the {4h@300°C} nanocomposite has a $M_r = 30.4(2)$ A m² kg⁻¹ and a $H_c = 90(1)$ kA m⁻¹, and these parameters are reduced by more than half for the sample with approximately the same composition fabricated at 400 °C for 2 h ($M_r = 13.8(2)$ A m² kg⁻¹, $H_c = 44.3(6)$ kA m⁻¹). Despite the similarity in composition between these two samples, the crystallite sizes of both hard and soft phases are much larger for the composite prepared at the higher temperature, which can explain the deterioration of the magnetic properties: (i) the 52.9(4) nm refined for the hard phase in {2h@400°C} is above the critical stable single-domain size (SSD) for CoFe₂O₄ ($\approx 40$ nm),[44] which explains the collapse in $H_c$ observed for this sample. (ii) The alloy also grows well beyond typical SSD values, and formation of domains in the soft phase eases spontaneous demagnetization of the hard when both phases are coupled.[31]

**Magnetization at low temperature (LT).** Magnetization *versus* applied field measured at 10 K is shown in Fig. 5(a) for selected samples: starting CoFe₂O₄ powders in green, {2h@300°C} in red, and {8h@300°C} in blue. The rest of the 10 K curves and the $M_s$, $M_r$, and $H_c$ values extracted may be found in Fig. S6 and Table S6 of the ESI,† respectively. LT magnetization measurements help understanding whether or not the hard and soft phases are linked through inter-particle exchange-coupling. Although the average reversal fields of CoFe₂O₄ and Co–Fe are similar at RT, they radically draw apart when lowering the temperature, as the anisotropy of the hard magnetic phase is significantly larger at LT.[45] This is clearly seen on our samples, with the $H_c$ of the hard phase being roughly 10 times larger at 10 K than at 300 K, while the $H_c$ of the soft phase {2h@600°C} is of the same order of magnitude at both temperatures (compare values from Table 2 and Table S6, ESI†). A discontinuous hysteresis loop is expected for uncoupled systems, as the hard and soft phases are independently demagnetized (two-step magnetization reversal). Oppositely, a smooth curve is expected for exchange-coupled systems, where a joint reversal of both phases takes place (1-step or single-phase reversal). The correlation single-/two-step LT hysteresis ↔ coupling/decoupling, respectively, is not always as simple as described above, but the statement is valid for the CoFe₂O₄/Co–Fe composite (see specific section in the ESI†).

The number of reversal or switching events is readily revealed by the maxima in the first derivative curve of the magnetization data. First derivatives of the M–H data from all

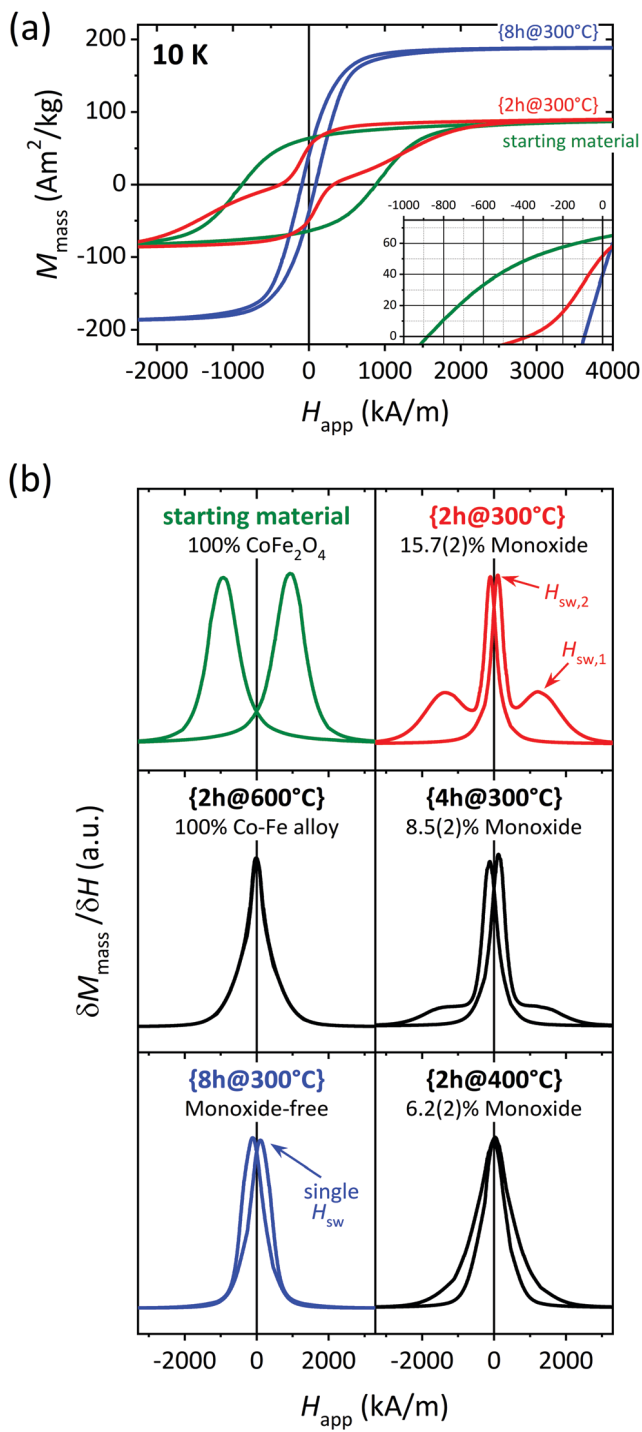

Fig. 5 (a) Low temperature magnetic hysteresis loops for selected samples and (b) corresponding first derivative normalized. The samples represented in green, red and blue color in this figure are highlighted with the same colors in Fig. 1 and 4.

convolution of several independent contributions from distinct phases (rather than a single-phase), all of them having a very similar, nearly-null magnetic anisotropy. This is in agreement with the two bcc species with different Co:Fe ratios visible in the PXRD data.

Two very distinct $H_{sw}$ are detected for {2h@300°C} (red), which is an indicative of weakly exchanged soft–hard interphases. On the contrary, {8h@300°C} (blue) presents a single-step reversal, which in this case is attributed to an effective of exchange-coupling between the soft and hard phases. Independent magnetization reversal of the magnetic phases is visible for {4h@300°C}, although the peak defined by the larger $H_{sw}$ is much less intense compared to the 2 h experiment at the same temperature (red curve). The $\delta M/\delta H$ curve for {2h@400°C} is maximized at a single $H_{sw}$ value. However, the peaks here are not symmetric and the peak tails do not coincide, suggesting some degree of decoupling of the two magnetic phases.

To summarize, the only composite showing LT exchange-coupling behavior is the monoxide-free sample {8h@300°C} (blue color). We believe this observation is far from coincidental, considering the correlation between the monoxide concentration and the degree of decoupling shown by our data (see plots on the right from Fig. 5(b)). The present study demonstrates how avoiding the monoxide is imperative for producing effectively exchange-coupled $CoFe_2O_4$/Co–Fe nanocomposites. This observation is consistent with and may help explain previous literature on the subject. Several studies report decoupling at RT in monoxide-containing samples[29,45–48] Some decoupled "monoxide-free" examples are also found.[49–53] However, we consider it possible that the monoxide was overlooked in those works. The proximity in $2\theta$ of the monoxide and the spinel Bragg positions (see Fig. 2) makes it difficult to separate the contribution from these two phases unless the PXRD data has high enough resolution and the subsequent data analysis is appropriate.

Based on a joint interpretation of the RT and LT magnetization data, we understand that our monoxide-free, exchange-coupled sample {8h@300°C} is far from reaching the best magnetic properties the $CoFe_2O_4$/Co–Fe system allows. Samples with a lower alloy content, such as {2h@300°C} and {4h@300°C}, appear significantly more promising, owing to their higher $M_r$ and $H_c$ values at RT (see Fig. 4), despite the presence of monoxide hindering the optimal magnetic performance of these samples. Therefore, we believe that monoxide-free composites with an alloy content $\lesssim 40$ wt% may lead to effectively exchange-coupled $CoFe_2O_4$/Co–Fe composites with superior magnetic properties, and should thus be pursued in future work.

## Conclusions

$CoFe_2O_4$ (hard)/Co–Fe (soft) magnetic nanocomposites have been prepared through controlled partial reduction of $CoFe_2O_4$ nanoparticles, obtaining samples with several compositions and crystallite sizes. An additional monoxide phase was found in some of the samples, although this phase disappeared for samples are displayed in Fig. 5(b). The starting material shows the single-step behavior expected for a pure phase, with a single switching field, $H_{sw}$, at $\approx 940$ kA m$^{-1}$. The same is observed for the fully-reduced sample {2h@600°C} but with a nearly zero $H_{sw}$. Note the shape of the peaks here is much more Lorentzian than for the starting material. This shape can result from the long reaction times. Magnetization curves at room and low temperature reveal that an increasing monoxide concentration deteriorates inter-phase magnetic exchange-coupling. In fact, the only composite showing an effective exchange-coupling was monoxide-free. Thus, minimizing/avoiding the formation of the monoxide is crucial for producing effectively exchange-coupled $CoFe_2O_4/Co$–Fe nanocomposites.

Once the chemistry behind the process is understood, partial reduction of $CoFe_2O_4$ is a very strong method for synthesizing $CoFe_2O_4/Co$–Fe nanocomposites with controlled magnetic properties. Adjusting each of the reduction parameters (temperature, time, partial $H_2$ pressure, crystallite size of the starting $CoFe_2O_4$ powders) has a very specific impact on the composition and crystallite sizes of the obtained nanocomposite, which, in turn, directly determines its magnetic behavior. The present work reveals exchange-coupling to be an excellent tool to further expand the range within which the magnetic properties of spinel ferrites can be tuned, extending the scope of this family of compounds. The method described here using $CoFe_2O_4/Co$–Fe as an example may in principle be applicable to other ferrite systems, including hard hexaferrites or other spinel ferrites (soft), and allows multiple combinations of magnetic compounds.

## Conflicts of interest

There are no conflicts to declare.

## Acknowledgements

C. G.-M. and A. Q. have contributed equally to this work. The authors would like to thank financial support from the European Commission through the AMPHIBIAN project (H2020-NMBP-2016-720853), the Danish National Research Foundation (Center for Materials Crystallography, DNRF-93), and the Spanish Ministerio de Ciencia, Innovación y Universidades (RTI2018-095303-A-C52). C. G.-M. acknowledges financial support from the Spanish Ministerio de Ciencia, Innovación y Universidades through the Juan de la Cierva Program (FJC2018-035532-I). Authors from Aarhus University gratefully acknowledge affiliation with the Center for Integrated Materials Research (iMAT) at Aarhus University. We acknowledge support of the publication fee by the CSIC Open Access Publication Support Initiative through its Unit of Information Resources for Research (URICI).

## Notes and references

1 D. L. Andrews, R. H. Lipson and T. Nann, *Comprehensive Nanoscience and Nanotechnology*, Elsevier B.V., 2019.
2 H. M. Williams, The application of magnetic nanoparticles in the treatment and monitoring of cancer and infectious diseases, *Biosci. Horizons Int. J. Student Res.*, 2017, **10**, 1–10.
3 Y.-L. Liu, D. Chen, P. Shang and D.-C. Yin, A review of magnet systems for targeted drug delivery, *J. Controlled Release*, 2019, **302**, 90–104.
4 P. Dong, K. P. Rakesh, H. M. Manukumar, Y. H. E. Mohammed, C. S. Karthik, S. Sumathi, P. Mallu and H. L. Qin, Innovative nano-carriers in anticancer drug delivery-a comprehensive review, *Bioorg. Chem.*, 2019, **85**, 325–336.
5 D. Chang, M. Lim, J. A. C. M. Goos, R. Qiao, Y. Y. Ng, F. M. Mansfeld, M. Jackson, T. P. Davis and M. Kavallaris, Biologically Targeted Magnetic Hyperthermia: Potential and Limitations, *Front. Pharmacol.*, 2018, **9**, 831.
6 K. Wu, D. Su, R. Saha, D. Wong and J.-P. Wang, Magnetic particle spectroscopy-based bioassays: methods, applications, advances, and future opportunities, *J. Phys. D: Appl. Phys.*, 2019, **52**, 173001.
7 T. A. P. Rocha-Santos, Sensors and biosensors based on magnetic nanoparticles, *TrAC, Trends Anal. Chem.*, 2014, **62**, 28–36.
8 G. Bülbül, A. Hayat and S. Andreescu, Portable Nanoparticle-Based Sensors for Food Safety Assessment, *Sensors*, 2015, **15**, 30736–30758.
9 Y. Li, Z. Wang, L. Sun, L. Liu, C. Xu and H. Kuang, Nanoparticle-based sensors for food contaminants, *TrAC, Trends Anal. Chem.*, 2019, **113**, 74–83.
10 D. Song, R. Yang, F. Long and A. Zhu, Applications of magnetic nanoparticles in surface-enhanced Raman scattering (SERS) detection of environmental pollutants, *J. Environ. Sci.*, 2019, **80**, 14–34.
11 M. Pastucha, Z. Farka, K. Lacina, Z. Mikušová and P. Skládal, Magnetic nanoparticles for smart electrochemical immunoassays: a review on recent developments, *Microchim. Acta*, 2019, **186**, 312.
12 J. M. Soares, V. B. Galdino and F. L. A. Machado, Exchange-bias and exchange-spring coupling in magnetic core–shell nanoparticles, *J. Magn. Magn. Mater.*, 2014, **350**, 69–72.
13 E. F. Kneller and R. Hawig, The exchange-spring magnet: a new material principle for permanent magnets, *IEEE Trans. Magn.*, 1991, **27**, 3588–3600.
14 G. Muscas, N. Yaacoub, G. Concas, F. Sayed, R. Sayed Hassan, J. M. Greneche, C. Cannas, A. Musinu, V. Foglietti, S. Casciardi, C. Sangregorio and D. Peddis, Evolution of the magnetic structure with chemical composition in spinel iron oxide nanoparticles, *Nanoscale*, 2015, **7**, 13576–13585.
15 H. L. Andersen, M. Saura-Múzquiz, C. Granados-Miralles, E. Canévet, N. Lock and M. Christensen, Crystalline and magnetic structure–property relationship in spinel ferrite nanoparticles, *Nanoscale*, 2018, **10**, 14902–14914.
16 H. L. Andersen, C. Granados-Miralles, M. Saura-Múzquiz, M. Stingaciu, J. Larsen, F. Søndergaard-Pedersen, J. V. Ahlburg, L. Keller, C. Frandsen and M. Christensen, Enhanced intrinsic saturation magnetization of $Zn_xCo_{1-x}Fe_2O_4$ nanocrystallites with metastable spinel inversion, *Mater. Chem. Front.*, 2019, **3**, 668–679.
17 T. Tatarchuk, M. Bououdina, J. J. Vijaya and L. J. Kennedy, in *Nanophysics, Nanomaterials, Interface Studies, and Applications*,

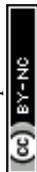

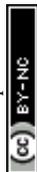

ed. O. Fesenko and L. Yatsenko, Springer International Publishing, 2017, pp. 305–325.

18 A. Hirohata, H. Sukegawa, H. Yanagihara, I. Zutic, T. Seki, S. Mizukami and R. Swaminathan, Roadmap for Emerging Materials for Spintronic Device Applications, *IEEE Trans. Magn.*, 2015, **51**, 1–11.

19 M. Amiri, M. Salavati-Niasari and A. Akbari, Magnetic nanocarriers: Evolution of spinel ferrites for medical applications, *Adv. Colloid Interface Sci.*, 2019, **265**, 29–44.

20 S. M. Hosseini, S. Sohrabnejad, G. Nabiyouni, E. Jashni, B. Van der Bruggen and A. Ahmadi, Magnetic cation exchange membrane incorporated with cobalt ferrite nanoparticles for chromium ions removal via electrodialysis, *J. Membr. Sci.*, 2019, **583**, 292–300.

21 Y. Wang, H. Li, L. Guo, Q. Jiang and F. Liu, A cobalt-doped iron oxide nanozyme as a highly active peroxidase for renal tumor catalytic therapy, *RSC Adv.*, 2019, **9**, 18815–18822.

22 C. Wei, Z. Feng, M. Baisariyev, L. Yu, L. Zeng, T. Wu, H. Zhao, Y. Huang, M. J. Bedzyk, T. Sritharan and Z. J. Xu, Valence Change Ability and Geometrical Occupation of Substitution Cations Determine the Pseudocapacitance of Spinel Ferrite $XFe_2O_4$ (X = Mn, Co, Ni, Fe), *Chem. Mater.*, 2016, **28**, 4129–4133.

23 Q. Zhao, Z. Yan, C. Chen and J. Chen, Spinels: Controlled Preparation, Oxygen Reduction/Evolution Reaction Application, and beyond, *Chem. Rev.*, 2017, **117**, 10121–10211.

24 F.-X. Ma, H. Hu, H. Bin Wu, C.-Y. Xu, Z. Xu, L. Zhen and X. W. David-Lou, Formation of Uniform $Fe_3O_4$ Hollow Spheres Organized by Ultrathin Nanosheets and Their Excellent Lithium Storage Properties, *Adv. Mater.*, 2015, **27**, 4097–4101.

25 Z. Feng, X. Chen, L. Qiao, A. L. Lipson, T. T. Fister, L. Zeng, C. Kim, T. Yi, N. Sa, D. L. Proffit, A. K. Burrell, J. Cabana, B. J. Ingram, M. D. Biegalski, M. J. Bedzyk and P. Fenter, Phase-Controlled Electrochemical Activity of Epitaxial Mg-Spinel Thin Films, *ACS Appl. Mater. Interfaces*, 2015, **7**, 28438–28443.

26 S. Liu, S. Sun and X.-Z. You, Inorganic nanostructured materials for high performance electrochemical supercapacitors, *Nanoscale*, 2014, **6**, 2037.

27 D. S. Mathew and R.-S. Juang, An overview of the structure and magnetism of spinel ferrite nanoparticles and their synthesis in microemulsions, *Chem. Eng. J.*, 2007, **129**, 51–65.

28 C. Granados-Miralles, M. Saura-Múzquiz, H. L. Andersen, A. Quesada, J. V. Ahlburg, A.-C. Dippel, E. Canévet and M. Christensen, Approaching Ferrite-Based Exchange-Coupled Nanocomposites as Permanent Magnets, *ACS Appl. Nano Mater.*, 2018, **1**, 3693–3704.

29 A. Quesada, F. Rubio-Marcos, J. F. Marco, F. J. Mompean, M. García-Hernández and J. F. Fernández, On the origin of remanence enhancement in exchange-uncoupled $CoFe_2O_4$-based composites, *Appl. Phys. Lett.*, 2014, **105**, 202405.

30 J. Ou-Yang, Y. Zhang, X. Luo, B. Yan, B. Zhu, X. Yang and S. Chen, Composition dependence of the magnetic properties of $CoFe_2O_4/CoFe_2$ composite nano-ceramics, *Ceram. Int.*, 2015, **41**, 3896–3900.

31 A. Quesada, G. Delgado, L. Pascual, A. M. Aragón, P. Marín, C. Granados-Miralles, M. Foerster, L. Aballe, J. E. Prieto, J. de la Figuera, J. F. Fernández and P. Prieto, Exchange-spring behavior below the exchange length in hard-soft bilayers in multidomain configurations, *Phys. Rev. B*, 2018, **98**, 214435.

32 J. V. Ahlburg, E. Canévet and M. Christensen, Air-heated solid–gas reaction setup for in situ neutron powder diffraction, *J. Appl. Crystallogr.*, 2019, **52**, 761–768.

33 Z. Guan, J. Jiang, N. Chen, Y. Gong and L. Zhen, Carbon-coated $CoFe_2$-$CoFe_2O_4$ composite particles with high and dual-band electromagnetic wave absorbing properties, *Nanotechnology*, 2018, **29**, 305604.

34 M. Lu, M. Liu, L. Wang, S. Xu, J. Zhao and H. Li, Structural and magnetic properties of $CoFe_2O_4/CoFe_2/SiO_2$ nanocomposites with exchange coupling behavior, *J. Alloys Compd.*, 2017, **690**, 27–30.

35 R. Safi, A. Ghasemi and R. Shoja-Razavi, The role of shell thickness on the exchange spring mechanism of cobalt ferrite/iron cobalt magnetic nanocomposites, *Ceram. Int.*, 2017, **43**, 617–624.

36 J. M. Soares, F. a. O. Cabral, J. H. de Araújo and F. L. A. Machado, Exchange-spring behavior in nanopowders of $CoFe_2O_4$-$CoFe_2$, *Appl. Phys. Lett.*, 2011, **98**, 072502.

37 A. Quesada, C. Granados-Miralles, A. López-Ortega, S. Erokhin, E. Lottini, J. Pedrosa, A. Bollero, A. M. Aragón, F. Rubio-Marcos, M. Stingaciu, G. Bertoni, C. de Julián Fernández, C. Sangregorio, J. F. Fernández, D. Berkov and M. Christensen, Energy Product Enhancement in Imperfectly Exchange-Coupled Nanocomposite Magnets, *Adv. Electron. Mater.*, 2016, **2**, 1500365.

38 M. Stingaciu, H. L. Andersen, C. Granados-Miralles, A. Mamakhel and M. Christensen, Magnetism in $CoFe_2O_4$ nanoparticles produced at sub- and near-supercritical conditions of water, *CrystEngComm*, 2017, **19**, 3986–3996.

39 J. Rodríguez-Carvajal, Recent advances in magnetic structure determination by neutron powder diffraction, *Phys. B*, 1993, **192**, 55–69.

40 P. Thompson, D. E. Cox and J. B. Hastings, Rietveld refinement of Debye–Scherrer synchrotron X-ray data from $Al_2O_3$, *J. Appl. Crystallogr.*, 1987, **20**, 79–83.

41 V. Lopez-Dominguez, A. Quesada, J. C. Guzmán-Mínguez, L. Moreno, M. Lere, J. Spottorno, F. Giacomone, J. F. Fernández, A. Hernando and M. A. García, A simple vibrating sample magnetometer for macroscopic samples, *Rev. Sci. Instrum.*, 2018, **89**, 034707.

42 I. Ohnuma, H. Enoki, O. Ikeda, R. Kainuma, H. Ohtani, B. Sundman and K. Ishida, Phase equilibria in the Fe–Co binary system, *Acta Mater.*, 2002, **50**, 379–393.

43 H. Zhang, D. Zeng and Z. Liu, The law of approach to saturation in ferromagnets originating from the magnetocrystalline anisotropy, *J. Magn. Magn. Mater.*, 2010, **322**, 2375–2380.

44 C. N. Chinnasamy, B. Jeyadevan, K. Shinoda, K. Tohji, D. J. Djayaprawira, M. Takahashi, R. J. Joseyphus and A. Narayanasamy, Unusually high coercivity and critical single-domain size of nearly monodispersed CoFe2O4 nanoparticles, *Appl. Phys. Lett.*, 2003, **83**, 2862–2864.
45 M. Kahnes, R. Müller and J. Töpfer, Phase formation and magnetic properties of $CoFe_2O_4/CoFe_2$ nanocomposites, *Mater. Chem. Phys.*, 2019, **227**, 83–89.
46 Y. Zhang, R. Xiong, Z. Yang, W. Yu, B. Zhu, S. Chen and X. Yang, Enhancement of Interparticle Exchange Coupling in $CoFe_2O_4/CoFe_2$ Composite Nanoceramics Via Spark Plasma Sintering Technology, *J. Am. Ceram. Soc.*, 2013, **96**, 3798–3804.
47 J. M. Soares, O. L. A. Conceição, F. L. A. Machado, A. Prakash, S. Radha and A. K. Nigam, Magnetic couplings in CoFe2O4/FeCo–FeO core–shell nanoparticles, *J. Magn. Magn. Mater.*, 2014, **374**, 1–5.
48 F. L. A. Machado, J. M. Soares, O. L. A. Conceição, E. S. Choi and L. Balicas, Magnetic properties of the nanocomposite $CoFe_2O_4$/FeCo-FeO at a high H/T regime, *J. Magn. Magn. Mater.*, 2017, **424**, 323–326.
49 Y. Zhang, Z. Yang, B. Zhu, S. Chen, X. Yang, R. Xiong and Y. Liu, Exchange-spring effect in $CoFe_2O_4/CoFe_2$ composite nano-particles, *J. Alloys Compd.*, 2013, **567**, 73–76.
50 X. Sun, Y. Q. Ma, Y. F. Xu, S. T. Xu, B. Q. Geng, Z. X. Dai and G. H. Zheng, Improved magnetic performance at low and high temperatures in non-exchange-coupling $CoFe_2O_4/CoFe_2$ nanocomposites, *J. Alloys Compd.*, 2015, **645**, 51–56.
51 J. Jin, X. Sun, M. Wang, Z. L. Ding and Y. Q. Ma, The magnetization reversal in $CoFe_2O_4/CoFe_2$ granular systems, *J. Nanopart. Res.*, 2016, **18**, 383.
52 G. Du and S. Wang, Synthesis of magnetically exchange coupled $CoFe_2O_4/CoFe_2$ core/shell composite particles through spray pyrolysis, *J. Alloys Compd.*, 2017, **708**, 600–604.
53 Y. Zhang, B. Yan, J. Ou-Yang, B. Zhu, S. Chen, X. Yang, Y. Liu and R. Xiong, Magnetic properties of core/shell-structured $CoFe_2/CoFe_2O_4$ composite nano-powders synthesized via oxidation reaction, *Ceram. Int.*, 2015, **41**, 11836–11843.